\documentclass[letter]{ptptex}

\usepackage{graphicx}
\usepackage{wrapft}
\usepackage{young}

\newcommand{\la}{\langle}
\newcommand{\ra}{\rangle}
\newcommand{\be}{\begin{equation}}
\newcommand{\ee}{\end{equation}}
\newcommand{\bea}{\begin{eqnarray}}
\newcommand{\eea}{\end{eqnarray}}

\title{
Heavy $qq$ interaction at finite temperature
}

\author{
Atsushi \textsc{Nakamura} and
Takuya \textsc{Saito}
}

\inst{
Research Institute for Information Science and Education,
Hiroshima University, Higashi-Hiroshima 739-8521, Japan
}

\abst{
The first lattice QCD numerical study of heavy quark-quark 
potentials at finite temperature is reported.
Using the quenched approximation, we evaluate the color anti-symmetric and
symmetric potentials.
}

\begin{document}
\maketitle

Recently, the diquark (a two-quark system) has attracted much
attention with regard to high-energy phenomenology.
Jaffe and Wilczek proposed that the recently discovered penta-quark
state ($\Theta^+$)\cite{Nakano} is a bound state of $(ud)(ud)\bar{s}$, where 
$(ud)$ represents for
 highly correlated $u$ and $d$ quark pairs.\cite{JaffeWilczek}
It is believed that at high baryon number density and low temperature  
there exists a family of color superconducting phases,
due to the quark pairing driven by the BCS mechanism.\cite{Alford}
( See Ref. \citen{DiquarkReview} for a review of the history of
diquarks and their role in high-energy reactions. )

A quark-quark system is a color anti-triplet (anti-symmetric) or
sextet (symmetric) state:
\bea
3 \times 3 &=& 3^* + 6\ , 
\\
\Young{1} \times \Young{1} &=& \Young[-1]{11} + \Young{2}\ .
\nonumber
\eea
We believe that
the quark-quark interaction is attractive and strong in the color 
anti-triplet channel, based on results obtained from    
the perturbation \cite{RGG} and instanton induced models \cite{instanton}.
It is important to investigate the quark-quark potential
using lattice QCD, which provides a non-perturbative and first principle
basis for exploring the quark-quark interaction.
To our knowledge, there has been only one such study, by Wetzorke
and collaborators\cite{Wetzorke}, in which diquark correlation
functions were calculated.


In this work, we study the heavy quark-quark free energies at finite
temperature that are obtained from the Polyakov line correlation (PLC).
PLC was first investigated by McLerran and Svetitsky\cite{McLerran}.
The free energy $F$ is given by
\be
 e^{-\beta F} = \sum_{\phi} \la\phi| e^{-\beta H} |\phi\ra,
\label{expF}
\ee
where $ |\phi\ra$ represents a state of gluons and heavy quarks.   
For the heavy quark-quark system of the color anti-triplet, we have 
\be
 |\phi\ra = \epsilon_{abc}\psi^b(\vec{x}_1,t=0)^\dagger
                          \psi^c(\vec{x}_2,t=0)^\dagger
                          |\mbox{Gluons}\ra.
\ee  
Here $a,b$ and $c$ are the color indices.
Now, the summation in Eq. (\ref{expF}) is taken over $a$ and all gluonic
states.
Using the relations\cite{McLerran} 
\begin{eqnarray}
\psi(\vec{x},t) = L \psi(\vec{x},0),
\\
\{ \psi^a(\vec{x},t=0),\psi^b(\vec{y},t=0)^\dagger \}
     = \delta_{a,b} \delta_{\vec{x},\vec{y}},
\end{eqnarray}
where
\be
 L(\vec{x}) = \prod_{t=1}^{N_t} U_{0}(\vec{x},t) 
\ee
and 
\be
  \epsilon_{abc}\epsilon_{ab'c'} = \delta_{b,b'}\delta_{c,c'}
                                 - \delta_{b,c'}\delta_{c,b'},
\ee
we obtain the free energies for the $qq$ sector 
in the symmetric and anti-symmetric channels,

\begin{equation}
\begin{array}{cll}
\displaystyle
\exp(-\beta {F_{s}(R)}) &=&\displaystyle
\frac{3}{4}{\la\mbox{Tr}L(\vec{x}_1)\mbox{Tr}L(\vec{x}_2)\ra}
+\frac{3}{4}{\la\mbox{Tr}L(\vec{x}_1)L(\vec{x}_2)\ra},\\
\displaystyle
\exp(-\beta{F_{as}(R)})&=&\displaystyle
\frac{3}{2}{\la\mbox{Tr}L(\vec{x}_1)\mbox{Tr}L(\vec{x}_2)\ra}
-\frac{3}{2}\la\mbox{Tr}L(\vec{x}_1)L(\vec{x}_2)\ra,\\
\end{array}\label{TrLL}
\end{equation}
where $F_s$ and $F_{as}$ are the differences in the free energy 
for the cases with and without
 the static color symmetric (anti-symmetric) quarks located at $\vec{x}_1$ and
$\vec{x}_2$, $R=|\vec{x}_1-\vec{x}_2|$ and $\beta=1/T$.


The formula (\ref{TrLL}) is gauge dependent,
 and therefore it requires the gauge fixing.
%
We employ the stochastic gauge fixing quantization (SGFQ)
with the Lorentz-type gauge fixing
proposed by Zwanziger\cite{Zwanziger}:
\begin{equation}
\frac{dA_{\mu}^a}{d\tau}=
-\frac{\delta S}{\delta A_{\mu}^a } +
\frac{1}{\alpha} D_{\mu}^{ab}(A)
\partial_{\nu} A_{\nu}^b + \eta_{\mu}^a\label{sq}.
\end{equation}
Here, $\alpha$, $D_{\mu}^{ab}$ and $\eta_{\mu}^a$
correspond to a gauge parameter,
a covariant derivative and Gaussian random noise, respectively.
The lattice formulation and a more detailed explanation
of this algorithm can be found in Refs. \citen{Mizutani,Saito,Saito2}.

In the quenched approximation, the system possesses $Z_3$ symmetry.
The gauge action is invariant under the transformation 
$U_4(x) \rightarrow z U_4(x)$ or $z^2 U_4(x)$ 
on a time-slice hyperplane, where $z = \exp(2\pi i/3)$.
In this transformation, 
$\la\mbox{Tr}L(\vec{x}_1)\mbox{Tr}L(\vec{x}_2)\ra$
and
$\la\mbox{Tr}L(\vec{x}_1)L(\vec{x}_2)\ra$ obtain a factor of $z^2$ or $z^4$.
After an infinite number of sweeps,
 the quantities in Eq. (\ref{TrLL}) vanish because $1+z^2+z^4=0$.
This is true also for the Polyakov line expectation value itself,
and often $ |\mbox{Tr}L| $ or $(\mbox{Re} (\mbox{Tr}L)^3)^{1/3}$ is taken.
The latter procedure brings the Polyakov lines around
$\pm 2\pi/3$ into regions around the positive real axis.
This may be justified because this area is realized in full QCD, 
due to the dynamical quark effect. 
Therefore we restrict our simulations to the region satisfying 
$-\pi/3 < \mbox{Phase } L < \pi/3$.


In order to calculate the color dependent Polyakov correlation functions
at finite temperature, we performed the quenched $SU(3)$ lattice simulation
with the standard plaquette action.
The values of the simulation parameters used 
are the same as in the previous study 
for the measurement of $q\bar{q}$ PLC functions in the QGP phase\cite{Saito3}.
The spatial lattice volume was $24^3$, and the temporal lattice 
size was set to $N_t=6$, which determines the temperature as $T=1/N_ta$,
where $a$ is the lattice spacing.
The corresponding critical temperature is
estimated to be $T_c \sim 256\ \mbox{MeV}$ in Ref. \citen{Boyd}.
The system temperature is changed when we vary the lattice cutoff, 
which is determined by Monte Carlo renormalization group analysis
 \cite{QCD_TARO}.
All of the PLCs are measured every ten Langevin steps and
normalized by the value $\la\mbox{Tr}L(0)\ra^2$.
We obtained between 3000 to 10000 data points 
after approximately 3000 steps were discarded as thermalization.

\begin{figure}[tbp]
\begin{center}
\resizebox{12cm}{!}
{\includegraphics{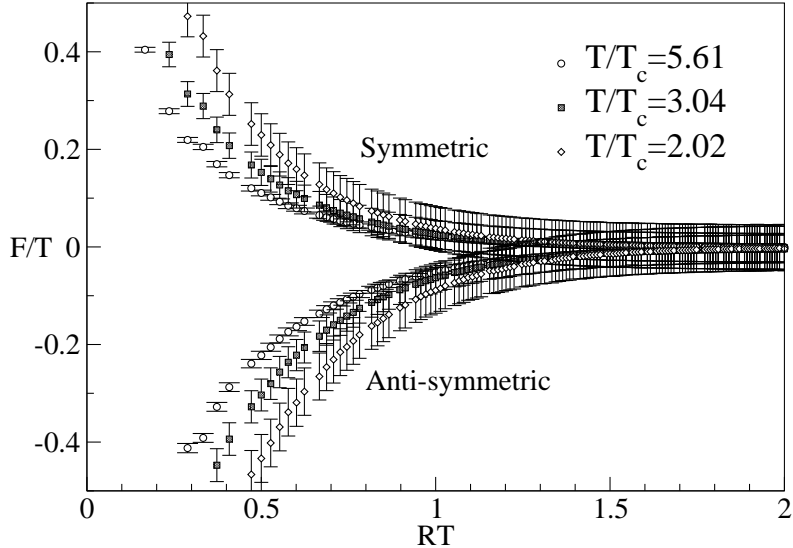}}
\caption{The behavior of qq free energies
 in the symmetric and anti-symmetric channels with Langevin step width
 $\Delta \tau = 0.03$.}\label{qq-zt}
\end{center}
\end{figure}

Typical behavior of the symmetric and anti-symmetric free energies 
at $T/T_c=2.02, 3.04, 5.61$ is displayed in Fig. \ref{qq-zt}.
The symmetric channel gives a repulsive force,
 while the anti-symmetric one gives an attractive force.
As the system temperature is varied, all potentials change but
their variations are observed to be small.

\begin{figure}[htbp]
\begin{center}
\resizebox{12cm}{!}
{\includegraphics{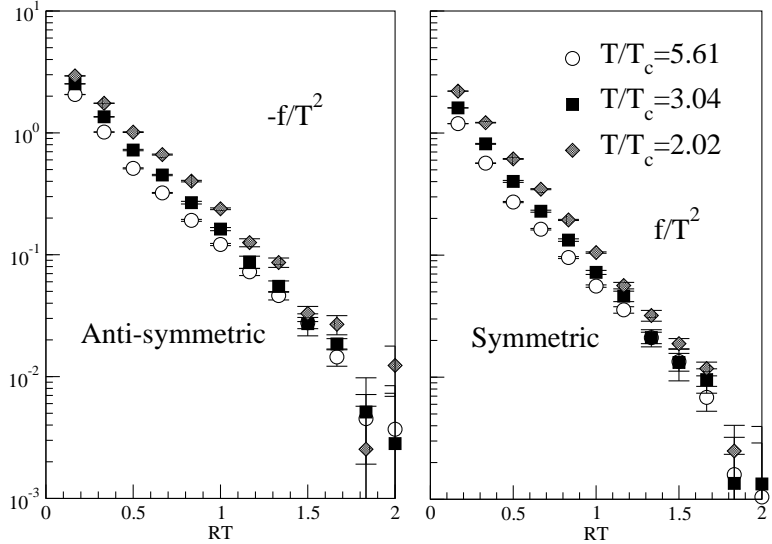}}
\caption{The temperature dependence of the effective force in the symmetric 
and anti-symmetric channels with Langevin step width $\Delta \tau = 0.03$.
}\label{force}
\end{center}
\end{figure}

If there is a strong attractive force between quarks even at high temperature,
diquark degrees of freedom should be taken into account for analyzing 
hadron production in high energy heavy ion collisions at RHIC and LHC.
Therefore it is important to investigate the diquark force by the 
lattice QCD simulation. 
As a first step for such a study, we define the effective force 
\begin{equation}
\frac{f(R,T)}{T^2} = -\frac{1}{T}(F(R+1)-F(R)).
\end{equation}
The temperature dependence of the strength of this effective force 
is shown in Fig. \ref{force}.
As the temperature increases, this strength decreases in both channels. 

\begin{figure}[htbp]
\begin{center}
\resizebox{12cm}{!}
{\includegraphics{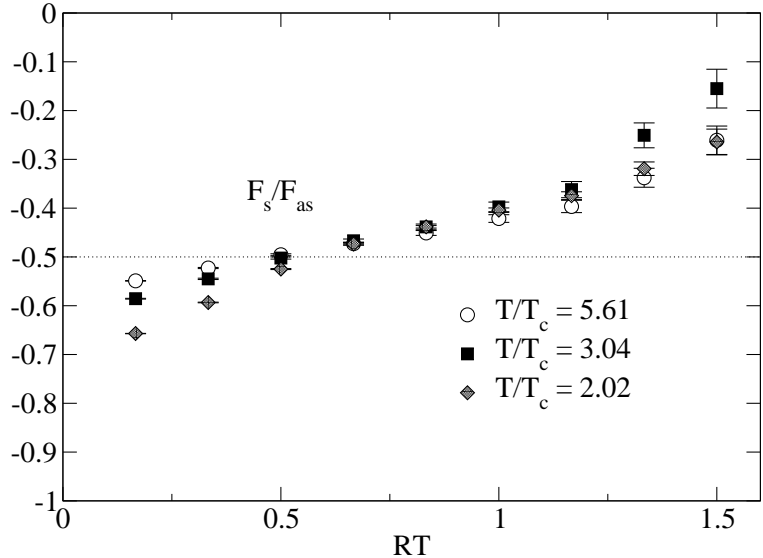}}
\caption{The temperature dependence of the ratio $F_{s}/F_{as}$
 with Langevin step width $\Delta \tau = 0.03$. The dash-dot line
stands for the expected ratio, $C_{qq}[6]/C_{qq}[3^{*}] = -0.5 $
at short distances. 
}\label{ratio}
\end{center}
\end{figure}

In the leading order perturbation (LOP), 
the coefficients of the color exchange terms
in the symmetric and anti-symmetric channels are 
$C_{qq}[6]=+\frac{1}{3}$ and $C_{qq}[3^{*}]=-\frac{2}{3}$ respectively.
We compare the force strengths in the symmetric and anti-symmetric channels
and define their ratio $F_{s}/F_{as}$, which approaches the 
value $C_{qq}[6]/C_{qq}[3^{*}] = -0.5 $ at short distances.
Figure \ref{ratio} displays values of this ratio evaluated 
 at $T/T_c=2.02, 3.04 \mbox{ and } 5.61$.
We find that as the temperature increases, three values at short distances 
become comparable with the expected values: However, as the temperature
decreases, it deviates from $-0.5$.
There are several possible sources for this deviation at shorter distances
 for small $T$.
One is numerical instability due to a long autocorrelation near $T_c$.
If this is indeed the cause, then larger scale simulations
 in the vicinity of $T_c$ are required.
Another possibility is the renormalization of PLC, i.e.,
 $\langle Tr L(R) L(0) \rangle = Z \exp(-F(R)/T)$.
If $Z$ differs significantly from 1, 
we should take it into account when comparing $F$
 with that obtained from the perturbation.
Recently, the Bielefeld group has proposed
 that the proper normalization of PLC
is determined by the comparison of the $T=0$ $q\bar{q}$ potential data with
the corresponding $T\neq0$ $q\bar{q}$ singlet free energy at short distances
\cite{karsch}.
However, the determination for the normalization of the $qq$ channels 
using this method is not performed.

\begin{figure}[hbt]
\begin{center}
\resizebox{12.0cm}{!}
{\includegraphics{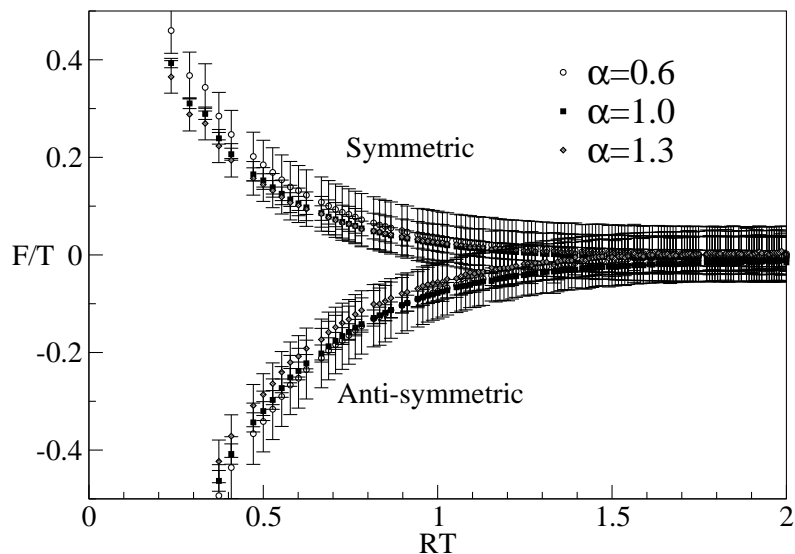}}
\end{center}
\caption{Gauge parameter $\alpha$ dependence of the free energies 
 in the symmetric 
and anti-symmetric channels  
 at $T/T_c=3.04$ ($\Delta \tau = 0.03$ ).
}\label{adep}
\end{figure}

In the SGFQ algorithm with a Lorentz-type gauge fixing term, 
 the $qq$ symmetric and anti-symmetric channels are studied 
by calculating the PLC functions. However, they are not gauge invariant.
Figure \ref{adep} displays the gauge parameter $\alpha$ dependence of 
$F_s$ and $F_{as}$ for three values of $\alpha$ from 0.6 to 1.3.
We find that the dependence on $\alpha$ in this range is small,
 and in particular, the basic features of the curves do not change.

In this paper we have reported the first non-perturbative lattice QCD
simulation of the $qq$ symmetric and anti-symmetric free energies.
Calculations were carried out for both gauge dependent channels 
in the framework of the stochastic quantization
 with a Lorentz-type gauge fixing term. 
We find that
 the anti-symmetric free energy constitutes an attractive force, 
while the symmetric free energy constitutes an repulsive force.  

\section*{Acknowledgements}
We thank Prof. M. Oka and Prof. D. Zwanziger for useful comments.
This work is supported by Grants-in-Aid for Scientific Research from
the Ministry of Education, Culture, Sports, Science and Technology,
Japan (No. 11694085, No. 11740159, and No. 12554008).

\end{document}